\documentclass{aa}
 \usepackage{graphicx} 
\begin{document}

   \title{Search for vertical stratification of metals in atmospheres of blue horizontal-branch stars}
   \author{V.R. Khalack \inst{1,}, F. LeBlanc\inst{1}, B.B. Behr\inst{2}\thanks{Current address:
US Naval Observatory, 3450 Massachusetts Avenue NW, Washington DC 20392}, G.A. Wade\inst{3}, D. Bohlender\inst{4}}

   \offprints{V. Khalack \\ \email{khalakv@umoncton.ca}}

   \institute{ D\'epartement de Physique et d'Astronomie,
               Universit\'e de Moncton,
               Moncton, N.-B., Canada E1A 3E9
               \and
               Department of Astronomy, University of Texas at Austin, 1 University
               Station C1400, Austin TX 78712-0259, USA
               \and
               Department of Physics, Royal Military College of Canada,
               PO Box 17000 stn `FORCES', Kingston, Ontario, Canada K7K 4B4
               \and
               National Research Council of Canada,
               Herzberg Institute of Astrophysics,
               5071 West Saanich Road, Victoria, BC, Canada V9E 2E7
             }

   \date{Received {\it date will be inserted by the editor}\
         Accepted {\it date will be inserted by the editor}
        }

\abstract 
{The observed abundance peculiarities of many chemical species relative to the expected cluster
metallicity in blue horizontal-branch (BHB) stars presumably appear as a result
of atomic diffusion in the photosphere.  The slow rotation (typically $v\sin{i}<$ 10 km s$^{-1}$)
of BHB stars with effective temperatures $T_{\rm eff}>$ 11,500~K supports this idea since the diffusion
mechanism is only effective in a stable stellar atmosphere.}
{In this work we search for observational evidence of vertical chemical
stratification in the atmospheres of six hot BHB stars: B84, B267 and B279 in M15
and WF2-2541, WF4-3085 and WF4-3485 in M13.}
{We undertake an abundance stratification analysis of the stellar atmospheres of
the aforementioned stars, based on acquired Keck HIRES spectra.}
{We have found from our numerical simulations that three stars (B267, B279 and
WF2-2541) show clear signatures of the vertical stratification of iron whose abundance increases toward
the lower atmosphere, while the other two stars (B84 and WF4-3485) do not. For WF4-3085 the
iron stratification results are inconclusive.
B267 also shows a signature of titanium stratification.
Our estimates for radial velocity, $v\sin{i}$ and overall iron, titanium and phosphorus abundances
agree with previously published data for these stars after taking the measurement errors into account.
The results support the hypothesis regarding the efficiency of atomic diffusion in the stellar atmospheres of
BHB stars with $T_{\rm eff}>$ 11,500~K.
}
{}
 \keywords{stars: atmospheres
        -- stars: horizontal-branch
        -- stars: chemically peculiar}

\titlerunning{Search for vertical stratification of metals in atmospheres of BHB stars}
\authorrunning{Khalack et al.}

\maketitle

\section{Introduction}

According to the current understanding of stellar evolution, the horizontal-branch (HB) stars
are post-main sequence stars that burn helium in their core
and hydrogen in a shell (e.g. Moehler \cite{Moehler04}). In this paper we consider the HB stars
that are located in the blue part of the HB, to the left of the RR Lyrae instability strip.
Most researchers call them blue horizontal-branch (BHB) stars to distinguish them from
the red horizontal-branch (RHB) stars, which exhibit different observational properties.
Sandage \& Wallerstein (\cite{S+W60}) have found from analysis of the colour-magnitude
diagrams of globular clusters\footnote{Most of the known BHB stars are found in globular clusters.}
that the HB 
generally becomes bluer with decreasing metallicity.
Derived masses of the cool ($T_{\rm eff}<$11,500~K) BHB stars in the globular cluster
NGC~6388 (Moehler \& Sweigart \cite{Moehler+S06}) are in a good agreement
with the predictions of
canonical HB evolution, except for the hot BHB stars with $T_{\rm eff}>$11,500~K, where
the estimated stellar masses seem to be lower than the canonical values.

\begin{table*}[th]
\parbox[t]{\textwidth}{
\centering
\caption[]{Journal of Keck+HIRES spectroscopic observations of the
selected hot BHB stars from Behr (\cite{Behr03b}).}
\begin{tabular}{lcccccc}
\hline
\hline
Cluster/Star&     HJD    &Exposure&  S/N & Coverage & Seeing & Slit' width\\
            &2450000+&   Time (s)   & &  (\AA)  & (arcsec) & (arcsec) \\
\hline
M13/WF2-2541& 1046.7902 &3$\times$1500& 44 & 3885-6292 & 0.90 & 0.86 \\
M13/WF4-3085& 1052.7338 &3$\times$1200& 37 & 3888-5356 & 1.10 & 0.86\\
M13/WF4-3485& 1053.7793 &3$\times$1200& 34 & 3888-5356 & 0.90 & 0.86\\
M15/B84     & 1053.9024 &4$\times$1400& 34 & 3888-5356 & 0.80 & 0.86\\
M15/B267    & 1053.9731 &4$\times$1400& 26 & 3888-5356 & 0.80 & 0.86\\
M15/B279    & 1053.8312 &4$\times$1400& 34 & 3888-5356 & 0.90 & 0.86\\
\hline
\end{tabular}
}
\label{tab1}
\end{table*}

The Hertzsprung-Russell diagrams of some globular clusters show long blue tails (an extension of the HB),
populated by very hot BHB stars and extreme horizontal branch (EHB) stars.
Published data on the BHB stars argue that the hot BHB stars
show remarkable differences in physical properties
when compared to the cool BHB stars.  Using high-precision
photometry of stars in M13, Ferraro et al. (\cite{Ferraro+98}) have found gaps in
the distribution of stars along the blue tail. One of these gaps, labeled as G1, is located at
$T_{\rm eff}\sim$11,000-12,000~K. Grundahl et al. (\cite{Grundahl+98}) used the results of
Str\"{o}mgren $uvby\beta$-photometry finding
good agreement between the theoretical prediction of
stellar evolution models and the observed location of BHB stars, except for the hot BHB stars,
whose $u$-magnitudes are brighter than predicted.
It appears that this $u$-jump is observed for the hot BHB stars
and coincides with the temperature range of the G1 gap in M13.  Similar
$u$-jumps have also been found for other globular clusters (Grundahl et al. \cite{Grundahl+99}).
For the hot BHB with effective temperatures up to 20,000 K, the surface gravities
derived from the fits of Balmer and helium line profiles appear to be lower than
the predictions of stellar evolution models,
while the gravities derived for the stars outside this temperature
range are in good agreement with theoretical predictions
(Moehler et al. \cite{Moehler+95, Moehler+97a, Moehler+97b, Moehler+03}).
The stellar rotation velocity distribution of BHB stars also appears to have a discontinuity
at $T_{\rm eff} \simeq $ 11,500~K (Peterson et al.~\cite{Peterson+95}; Behr et al. \cite{Behr+00a};
Recio-Blanco et al.~\cite{RB+04}), indicating that the hotter stars show modest rotation
with $v\sin{i}<$ 10 km s$^{-1}$, while the cooler stars rotate more rapidly on average.
Comprehensive surveys of abundances also show that
the hot BHB stars
have abundance anomalies when compared to the cool BHB stars in the same
cluster (Glaspey et al. \cite{Glaspey+89}; Grundahl et al. \cite{Grundahl+99};
Behr et al. \cite{Behr+99, Behr+00b}, Behr \cite{Behr03a}; Fabbian et al. \cite{Fabbian+05};
Pace et al. \cite{Pace+06}).

The observed phenomena such as the low gravity, photometric jumps and gaps, abundance anomalies
and slow rotation suggest that atomic diffusion could be important in the stellar atmospheres of
hot BHB stars. Atomic diffusion arises from
the competition between radiative acceleration and gravitational settling.
This can produce a net acceleration on atoms and ions,
which results in their diffusion through the atmosphere (Michaud~\cite{Michaud70}).
In order for atomic diffusion to produce a vertical stratification of the abundances of
particular elements, the stellar atmosphere must be hydrodynamically stable.
According to Landstreet (\cite{Landstreet98}), photospheric convection should be very
weak at the effective temperatures of BHB stars. Theoretical atmospheric models
of Hui-Bon-Hoa, LeBlanc \& Hauschildt (\cite{Hui-Bon-Hoa+00}) showed that the observed
photometric jumps and gaps for hot BHB stars can be explained by elemental diffusion
in their atmosphere. Behr (\cite{Behr03b}) has shown that adoption
of a microturbulent velocity of 0 or 1 km s$^{-1}$ provides the best fit to line
strengths in the spectra of hot BHB stars. This fact supports the proposal that
strong velocity fields are not present in the atmospheres of hot BHB stars.

While synthesizing spectral line profiles, Khalack et al. (\cite{Khalack+07}) have recently
found vertical abundance stratification of sulfur in the atmosphere of the field BHB star HD~135485.
In this paper we also attempt to detect signatures of vertical abundance stratification
of elements from line profile analyses of several other BHB stars for which we have appropriate
data. Together with the data
on stratification of the sulfur abundance in HD~135485, new positive results
would provide a convincing argument in favour of efficient atomic diffusion
in the atmospheres of hot BHB stars.
In Sec.~\ref{obs} we discuss the properties of the acquired spectra, while in Sec.~\ref{mod}
we describe details concerning the simulation routine and adopted atmospheric parameters
for the program stars. The evidence for vertical stratification of some chemical species
is given in Sec.~\ref{vert}, while the estimation of mean abundances and velocities is
described in Sec.~\ref{mean}. A discussion follows in Sec.~\ref{discuss}.

\section{Observations}
\label{obs}

In this paper we have selected hot
BHB stars from the list of objects
published by Behr (\cite{Behr03b}) and
which have comparatively high signal-to-noise ratio (S/N) spectra available.
Spectroscopic observations of the selected stars were undertaken
in August 1998 with the Keck I telescope and the HIRES spectrograph.
The journal of spectroscopic observations is shown in
Table~\ref{tab1} where individual columns give the object identification,
the heliocentric Julian Date of the observation, the exposure time, the S/N per pixel
the spectral coverage, the size of the seeing disk (FWHM) and the size of the C1 slit that
provides a spectral resolution of $R=\lambda/\delta\lambda=$45,000.
For the aforementioned stars, Behr (\cite{Behr03b}) has found that the underfilling of the slit
should not change the estimated spectral resolution by more than 4\%--7\%.

The package of routines developed by McCarthy (\cite{McCarthy90}) for the FIGARO
data analysis package (Shortridge~\cite{Shortridge93}) was employed  to process the spectra.
A comprehensive description of the data acquisition and reduction procedure is presented
by Behr~(\cite{Behr03b}). This followed a standard
prescription with bias subtraction, flat-fielding, order extraction,
and wavelength calibration from thorium-argon comparison lamp observations.
To minimize the potential distortion of narrow spectra features, cosmic
ray hits were identified and removed by hand.

\section{Line profile simulations}
\label{mod}

\subsection{Stellar atmosphere parameters}

The line profile simulations were performed with the {\sc Zeeman2} spectrum
synthesis code (Landstreet~\cite{Landstreet88}; Wade~et~al.~\cite{Wade+01}).
The stellar atmosphere models were calculated with the {\sc Phoenix} code
(Hauschildt et al.~\cite{Hauschildt+97}) assuming LTE (Local Thermodynamic Equilibrium)
and using the stellar atmosphere parameters extracted from Behr (\cite{Behr03b})
and listed in Table~\ref{tab2}. In calculating the
stellar atmosphere models for program stars, we have used solar metallicity
with the enhanced iron and depleted helium abundances derived by Behr (\cite{Behr03b}).
The depleted helium abundance has also been taken into account during
line profile simulations using {\sc Zeeman2}.
We used gaussian instrumental profiles with widths derived from the comparison arc spectra.

\begin{figure*}[th]
\parbox[t]{\textwidth}{
\centerline{%
\begin{tabular}{@{\hspace{+0.05in}}c@{\hspace{+0.1in}}c}
a) & b)\\
\includegraphics[angle=-90,width=3.5in]{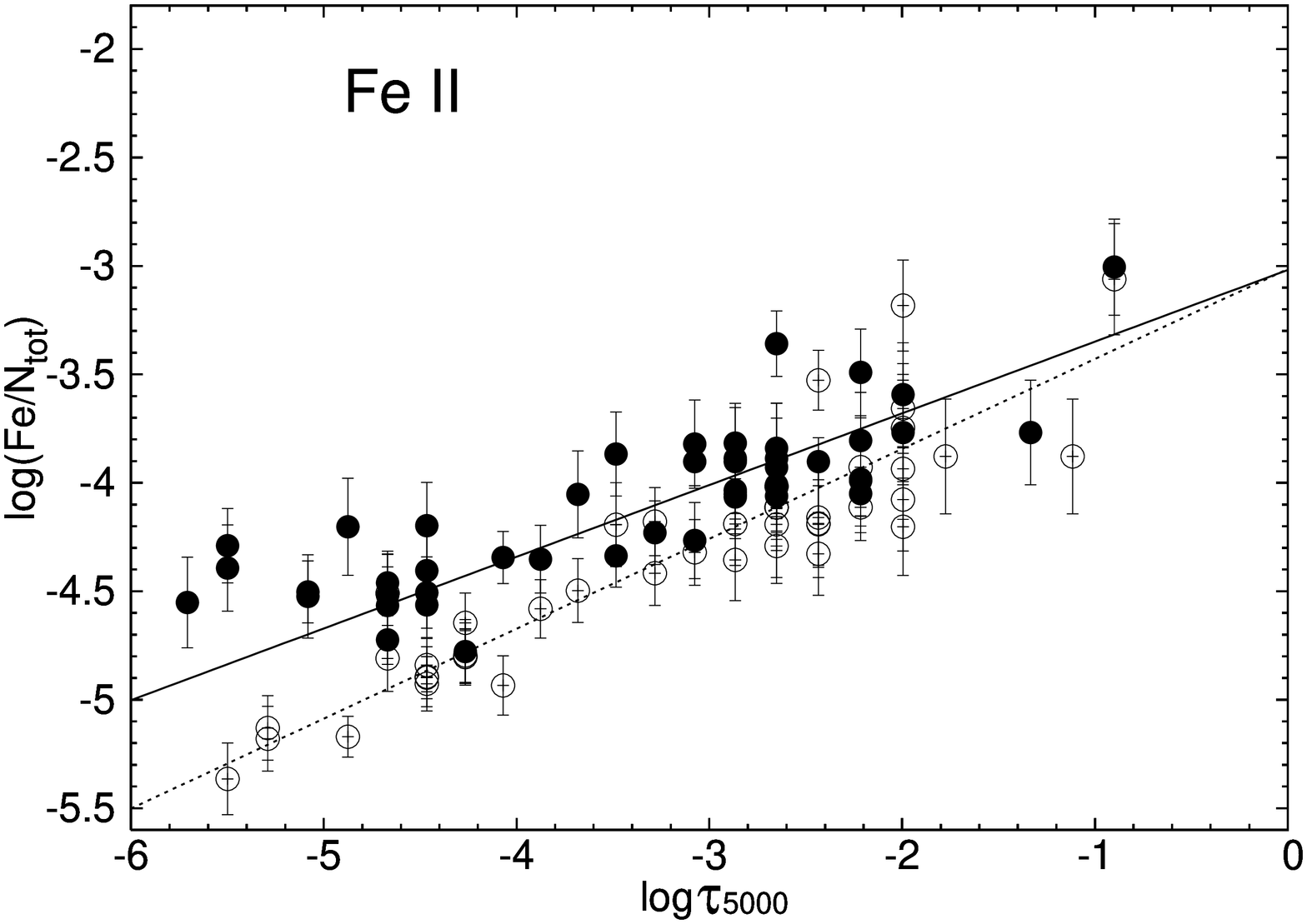} &
\includegraphics[angle=-90,width=3.5in]{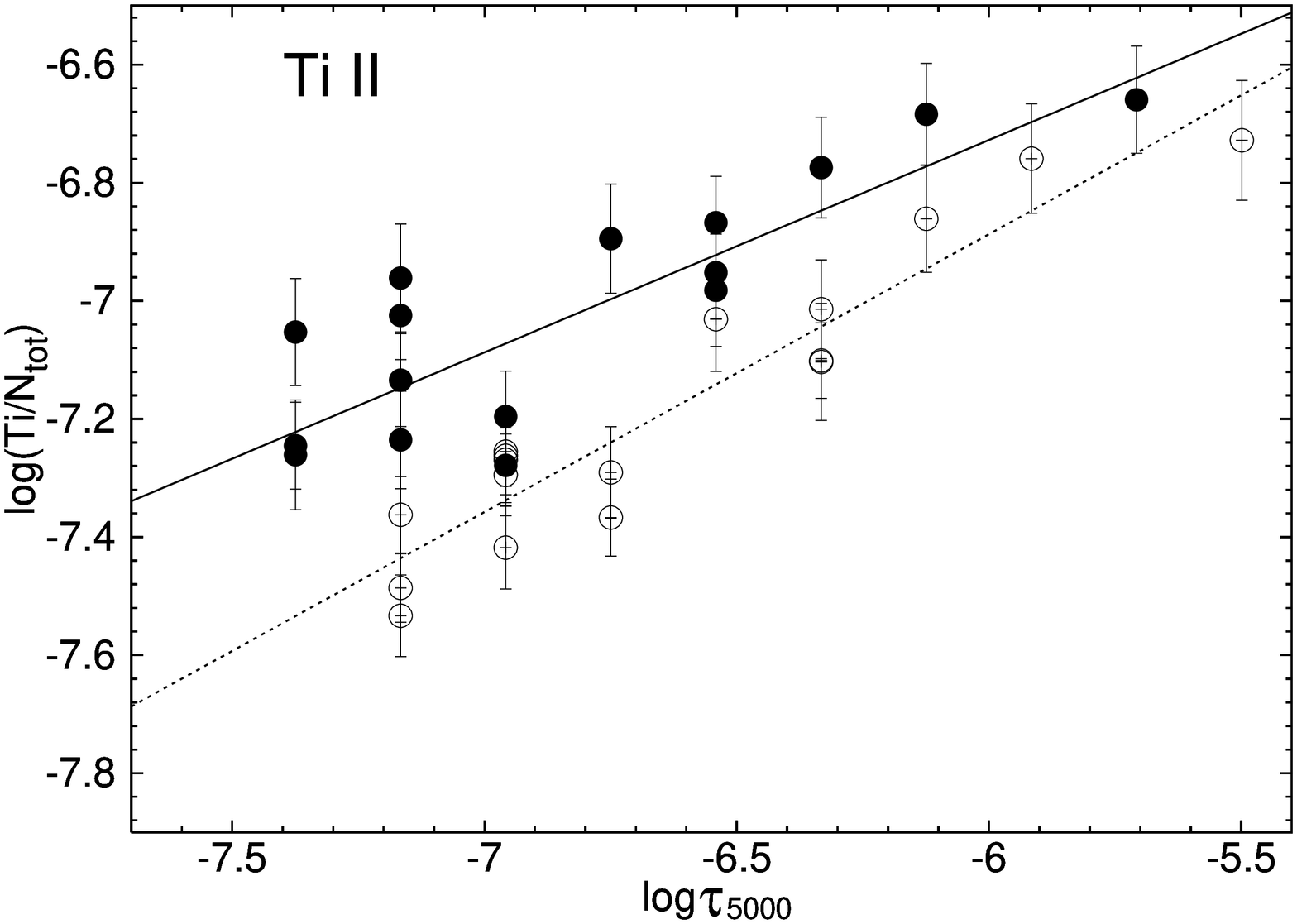}
\end{tabular}
}
\caption{Abundance estimates from the analysis of (a) Fe\,{\sc ii} and (b) Ti\,{\sc ii}
lines as a function of line (core) formation optical depth assuming $\xi$=0 km s$^{-1}$
(filled circles) and $\xi$=2 km s$^{-1}$ (open circles) for B267. The solid and dashed lines
approximate the filled and open circles respectively with straight line using the least-square
algorithm.
\label{B267}}
}
\end{figure*}

\begin{figure*}[th]
\parbox[t]{\textwidth}{
\centerline{%
\begin{tabular}{@{\hspace{+0.05in}}c@{\hspace{+0.1in}}c}
a) & b)\\
\includegraphics[angle=-90,width=3.5in]{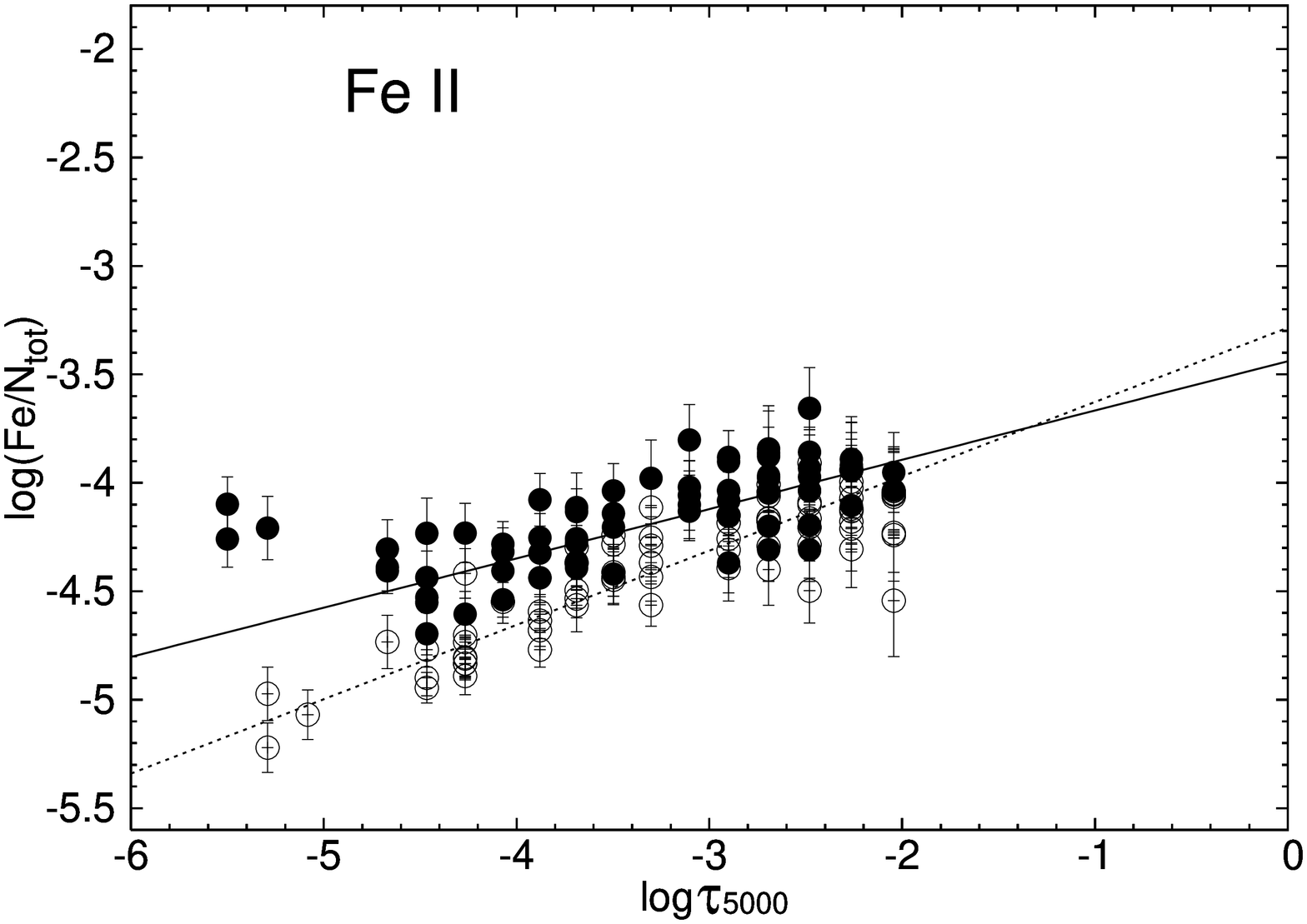} &
\includegraphics[angle=-90,width=3.5in]{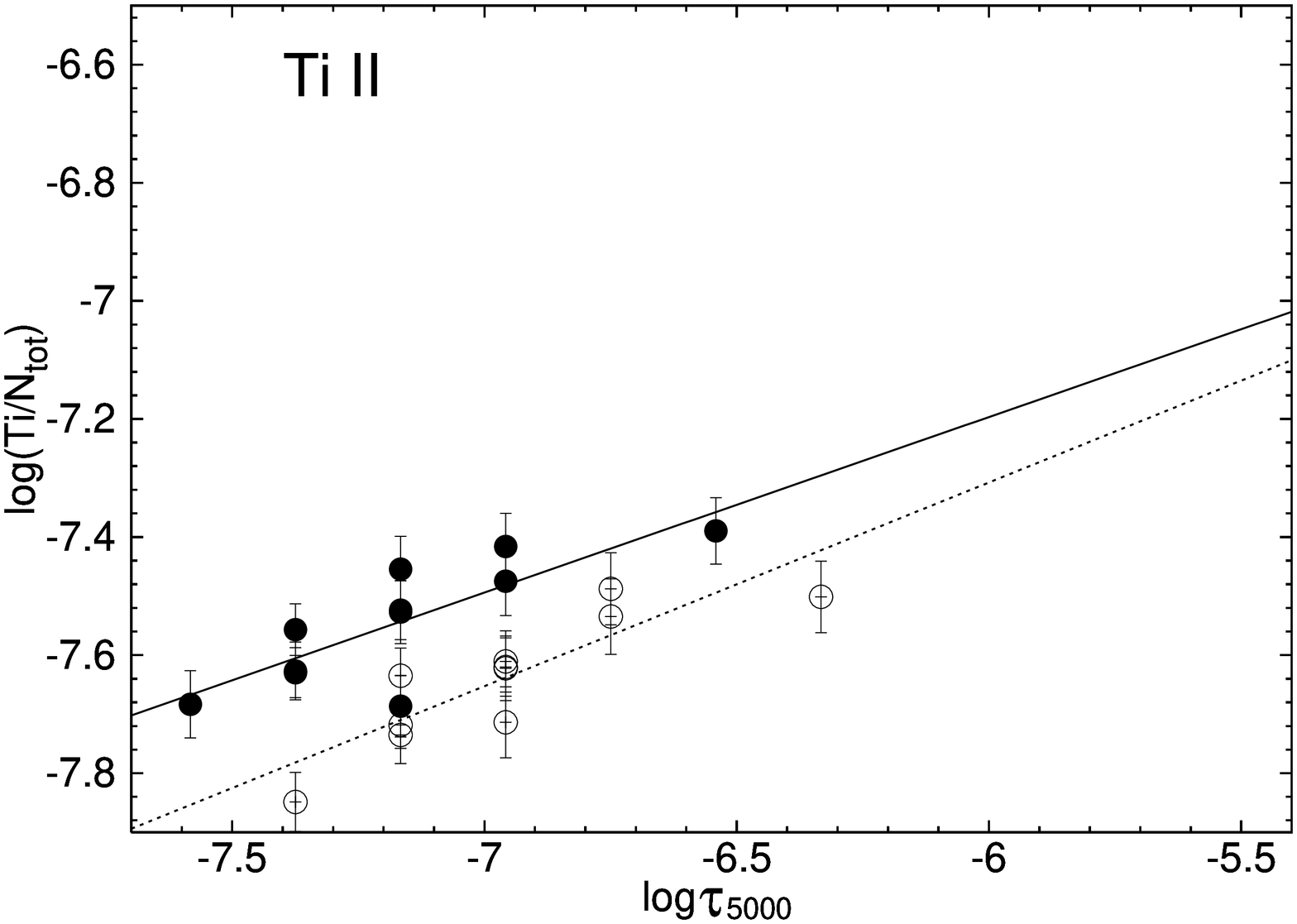}
\end{tabular}
}
\caption{The same as in Fig.~\ref{B267}, but for (a) Fe\,{\sc ii} and (b) Ti\,{\sc ii}
lines extracted from the B279 spectrum.
\label{B279}}
}
\end{figure*}

To simulate the spectra of BHB stars, Behr (\cite{Behr03b}) initially adopted a microturbulent
velocity $\xi=$2 km s$^{-1}$ and subsequently updated it during the simulation routine according to the results
of his model fits. He has obtained best fit values of $\xi$ from 0 to 0.9 km s$^{-1}$ for the BHB stars studied
here (see Table~\ref{tab2}).
Since we search for possible vertical abundance
stratification, our results can be affected by microturbulence. To estimate this influence we have adopted
here two different microturbulent velocities $\xi=$0 and 2 km s$^{-1}$ for our simulations.

\begin{table}[t]
\caption[]{Parameters of stellar atmospheres for the selected hot BHB stars from Behr (\cite{Behr03b}).}
\begin{tabular}{lcccc}
\hline
\hline
Cluster/Star&$T_{\rm eff}$&$\log{g}$&   $\xi$  &$V_{\rm r}$\\
            &     (K)        &  (dex)      &(km s$^{-1}$)&(km s$^{-1}$) \\
\hline
M13/WF2-2541& 13000 & 4.0 & 0.0 &-257.5  \\
M13/WF4-3085& 14000 & 4.0 & 0.0 &-255.7  \\
M13/WF4-3485& 13000 & 4.0 & 0.0 &-246.8  \\
M15/B84     & 12000 & 3.5 & 0.5 &-108.2  \\
M15/B267    & 11000 & 3.5 & 0.0 &-114.4  \\
M15/B279    & 11000 & 3.5 & 0.9 &-104.4  \\
\hline
\end{tabular}
\label{tab2}
\end{table}

\subsection{Procedure}
\label{proc}

We examined the spectrum of each star to establish a line list
suitable for abundance and stratification analysis. The line
identification was performed using the VALD-2
(Kupka~et~al.~\cite{Kupka+99}; Ryabchikova~et~al.~\cite{Ryab+99}),
Pickering et al. (\cite{Pickering+01}) and Raassen \&
Uylings\footnote{\rm ftp://ftp.wins.uva.nl/pub/orth} (\cite{RU+98})
line databases. For all of the program stars the best represented
element (with a number of readily visible line profiles) is iron.
Some stars also show Ti\,{\sc ii} or P\,{\sc ii}
lines. Therefore iron, titanium and phosphorus
were selected for analysis of their possible vertical abundance stratification.
For Fe\,{\sc ii} lines, we extracted atomic data from Raassen \& Uylings (\cite{RU+98}),
while for P\,{\sc ii} lines we used VALD-2.
Atomic data for Ti\,{\sc ii} lines were taken from Pickering et al. (\cite{Pickering+01}).

\begin{figure*}[th]
\parbox[t]{\textwidth}{
\centerline{%
\begin{tabular}{@{\hspace{+0.05in}}c@{\hspace{+0.1in}}c}
a) & b)\\
\includegraphics[angle=-90,width=3.5in]{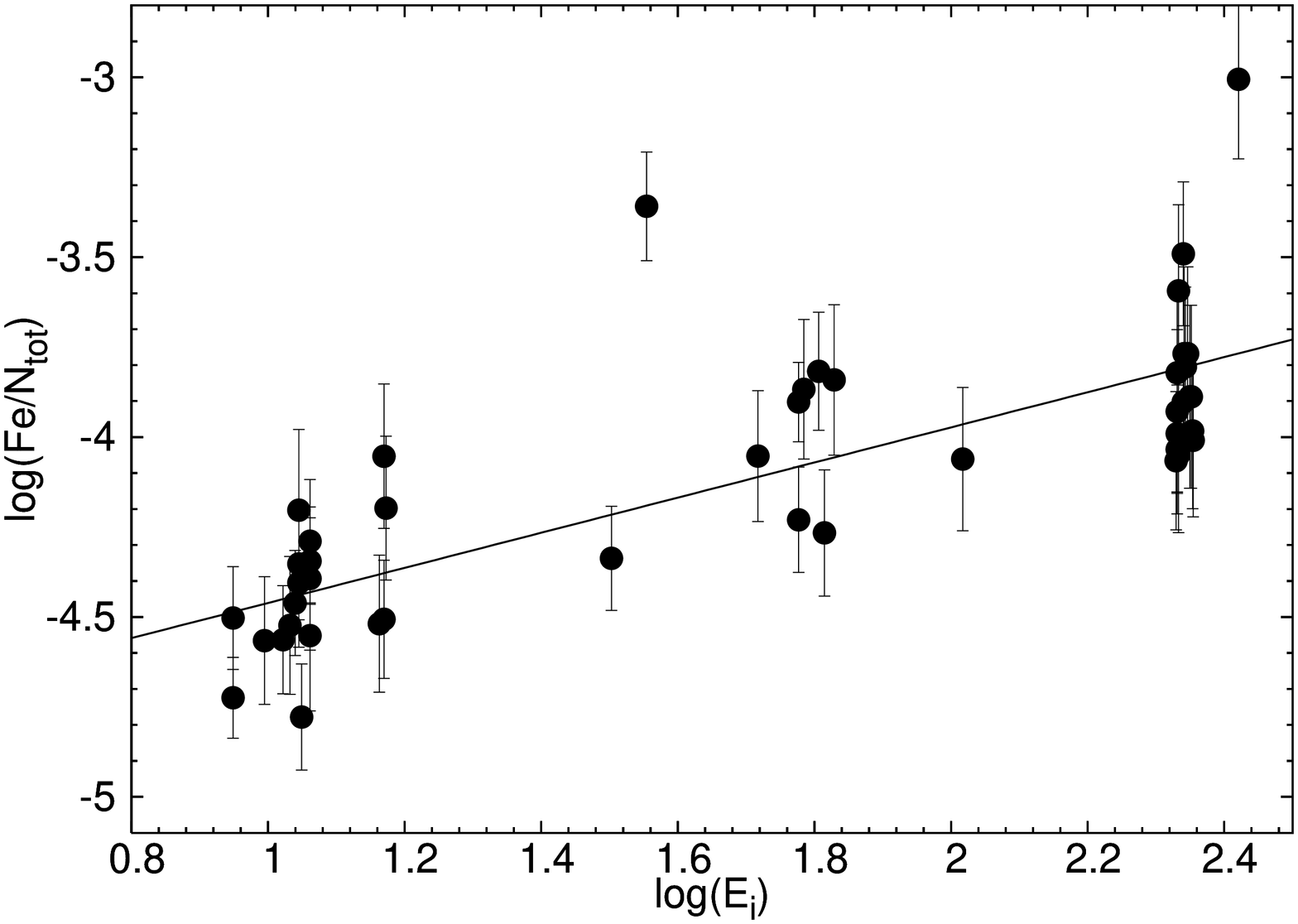} &
\includegraphics[angle=-90,width=3.5in]{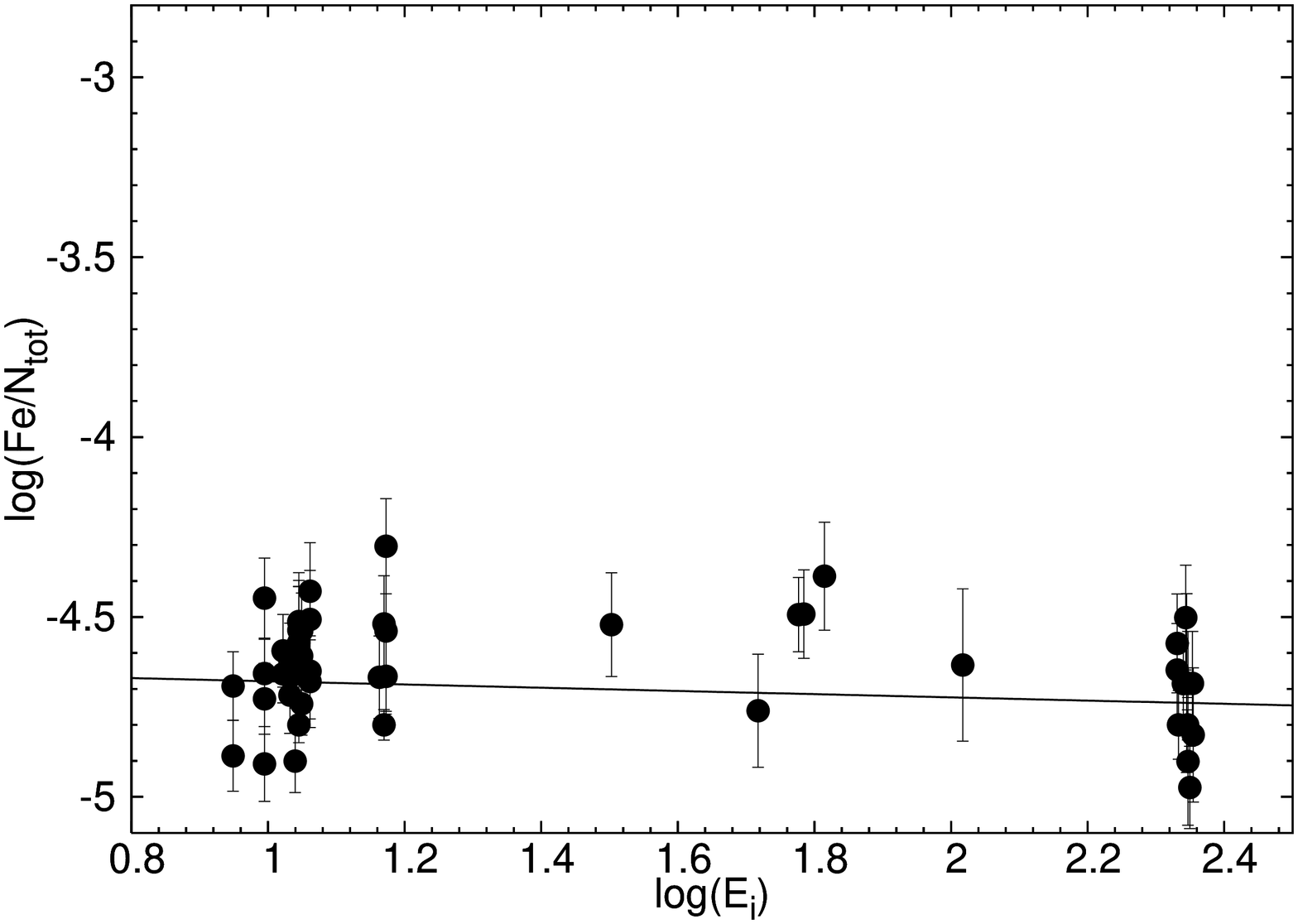}
\end{tabular}
}
\caption{Dependence of iron abundance on the lower level excitation potential
of Fe\,{\sc ii} lines extracted from (a) B267 and (b) WF4-3485 spectra.
The linear least-square fit to the data
is shown by solid lines. Here the star B267 shows direct evidence of iron stratification,
while WF4-3485 does not.
\label{excitation}}
}
\end{figure*}

\begin{table*}[t]
\centering
\caption[]{List of iron spectral lines used for the abundance analysis.}
\begin{tabular}{crccl|cccccc}
\hline \hline
$\lambda$, \AA&$\log gf$& $E_i, cm^{-1}$&$\log \gamma_{\rm rad}$& Ref.&\multicolumn{5}{c}{ Object name}\\
\hline
\multicolumn{5}{c|}{ Fe\,{\sc ii}}    &WF2-2541&WF4-3085&WF2-3585&B84&B267&B279\\
\hline
4122.668&-3.300& 20830.58& 8.49 & R\&U &  ...   &   x   &  ...   & x & ...& ...\\
4128.748&-3.578& 20830.58& 8.61 & R\&U &  ...   &   x   &  ...   &...& ...&  x \\
4173.461&-2.617& 20830.58& 8.61 & R\&U &   x    &   x   &    x   & x &  x & ...\\
4177.618&-3.776& 73395.93& 8.61 & R\&U &  ...   &   x   &  ...   & x & ...&  x \\
4177.692&-3.449& 20516.96& 8.61 & R\&U &  ...   &   x   &  ...   & x & ...&  x \\
4178.862&-2.535& 20830.58& 8.49 & R\&U &   x    &   x   &    x   & x & ...&  x \\
4184.261&-1.938& 90397.87& 8.35 & R\&U &  ...   &   x   &  ...   &...& ...& ...\\
\hline
\end{tabular}
\label{tab3}
\note{Table~\ref{tab3} is presented in full in the electronic edition of Astronomy and Astrophysics.
A portion is shown here for guidance and content.}

\end{table*}

The {\sc Zeeman2} code has been modified (Khalack~\&~Wade \cite{Khalack+Wade06}) to
allow for an automatic minimization of the model parameters using the
{\it downhill simplex method} (Press~et~al.~\cite{press+}). The minimization
routine finds the global minimum of the $\chi^\mathrm{2}$ function, which is specified
as the measure of differences between simulated and observed line profiles.
The relatively poor efficiency of the downhill simplex method, requiring a
large number of function evaluations, is a well known problem.
However, repeating the minimization routine 
several times in the vicinity of a supposed minimum in the chosen parameter space allows
us to verify if the method converges to the global minimum (for more details about this
minimization routine see Khalack et al. \cite{Khalack+07}).

To search for the presence of abundance stratification, we have
estimated the abundance of a chemical element
from an independent analysis of each selected line profile.
This method operates with three free model parameters (the element's abundance,
the radial velocity $V_{\rm r}$ and $v\sin{i}$) that are derived from each line profile
using the aforementioned automatic minimization routine. To analyse the vertical abundance
stratification in this method we build the scale of optical depths $\tau_{\rm 5000}$
for the list of selected line profiles. First, we calculate the line optical depth $\tau_{\rm \ell}$
in the line core for every layer of the stellar atmosphere model.
Next, we suppose that each analyzed profile is formed mainly at $\tau_{\rm \ell}$=1,
which corresponds to a particular layer of the stellar atmosphere with respective continuum
optical depth $\tau_{\rm 5000}$.
All simulations are performed with stellar atmosphere models that contain 50 layers.

In general, we have selected for our analysis lines found to be free of predicted or inferred blends.
However, if a blend is from a line of the same element that forms the main line profile,
such a line was also included in our simulation.
When a simulated line profile results in a radial velocity which differs significantly
(more than 2 km s$^{-1}$) from the average $V_{\rm r}$ for the analysed star,
we exclude such a line from further consideration. The difference in radial velocity may
be evidence of line misidentifications or inaccurate line wavelengths.
Atomic data (and sources) for the final list
of iron lines selected for our study is given in Table~\ref{tab3}.

\begin{figure*}[th]
\parbox[t]{\textwidth}{
\centerline{%
\begin{tabular}{@{\hspace{+0.05in}}c@{\hspace{+0.1in}}c}
a) & b)\\
\includegraphics[angle=-90,width=3.5in]{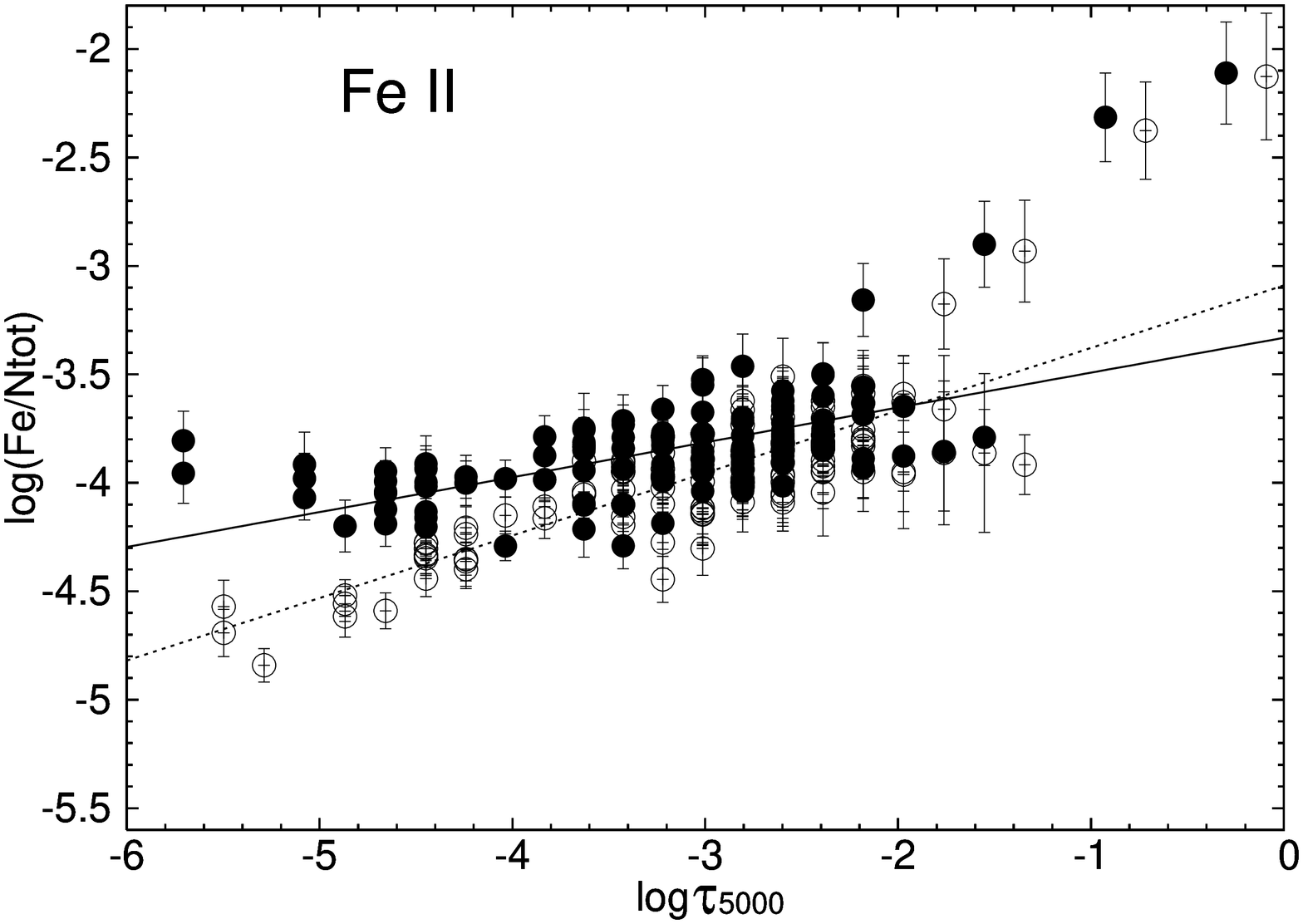} &
\includegraphics[angle=-90,width=3.5in]{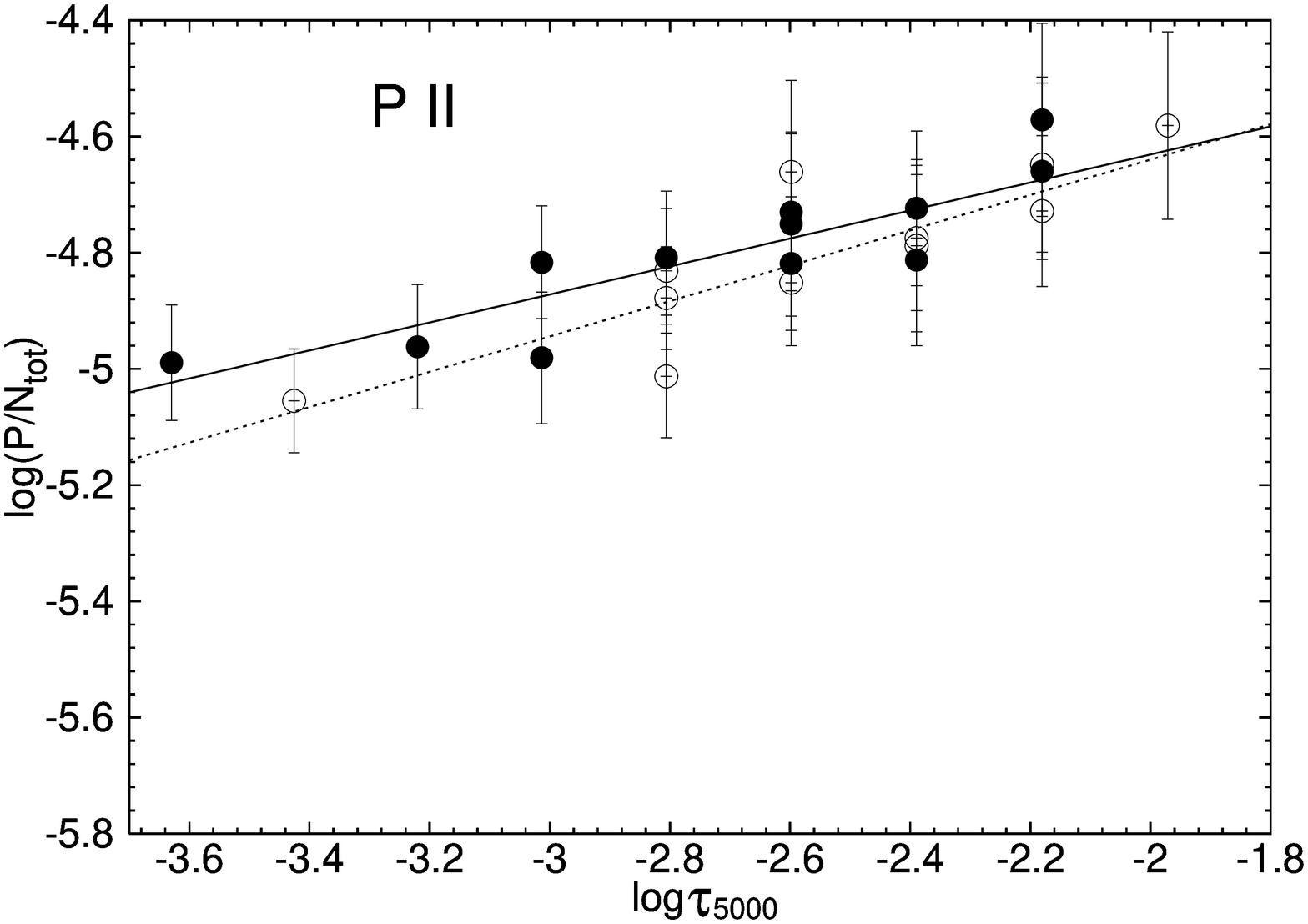}
\end{tabular}
}
\caption{The same as in Fig.~\ref{B267}, but for (a) Fe\,{\sc ii} and (b) P\,{\sc ii}
lines extracted from the WF4-3085 spectrum. Here the visual strong rise in iron
abundance at large optical depth manifests itself by four lines and may be real.
Nevertheless, to draw the final conclusion we need more data for this range of optical depths.
\label{WF4-3085}}
}
\end{figure*}

\begin{figure*}[th]
\parbox[t]{\textwidth}{
\centerline{%
\begin{tabular}{@{\hspace{+0.05in}}c@{\hspace{+0.1in}}c}
a) & b)\\
\includegraphics[angle=-90,width=3.5in]{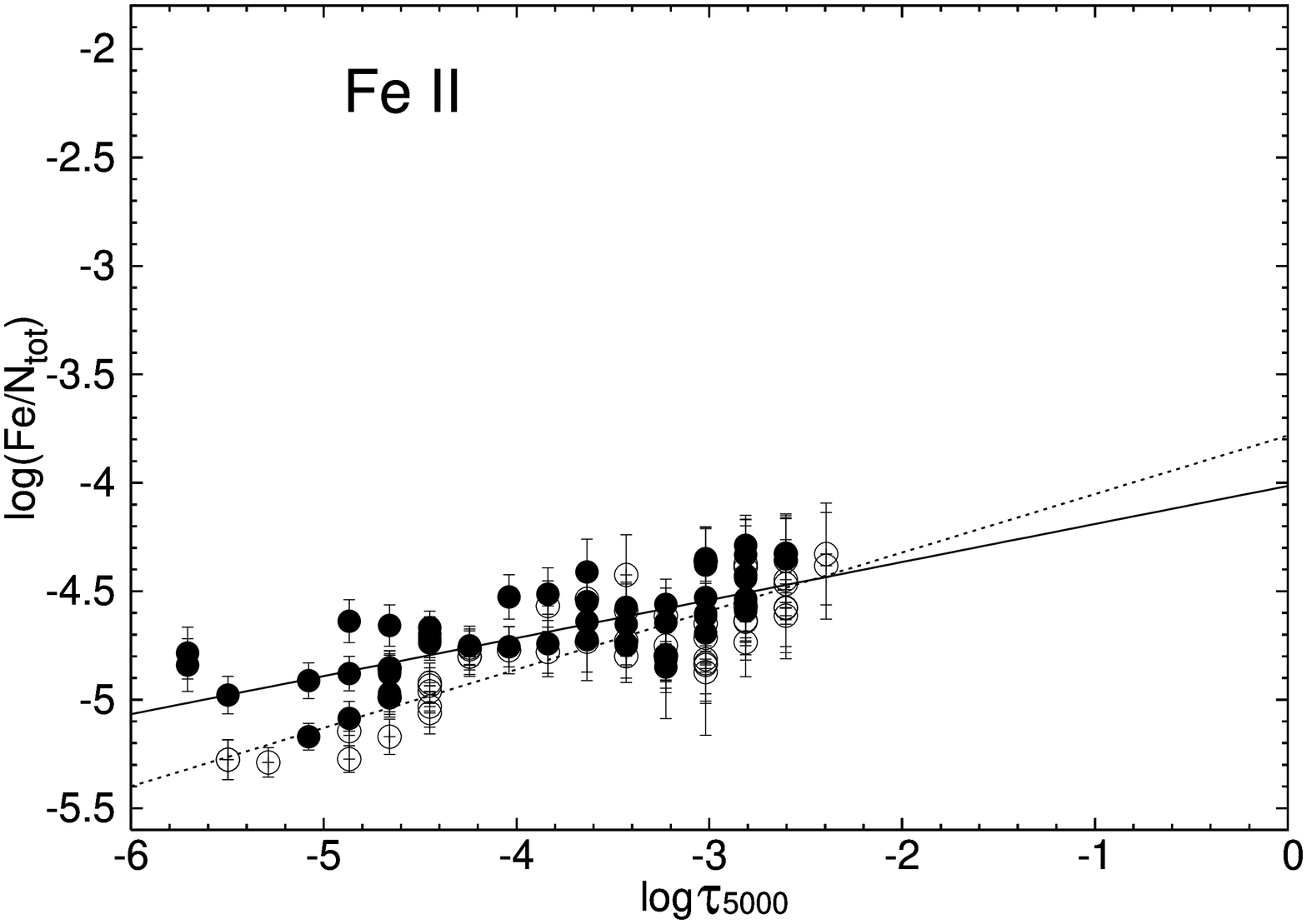} &
\includegraphics[angle=-90,width=3.5in]{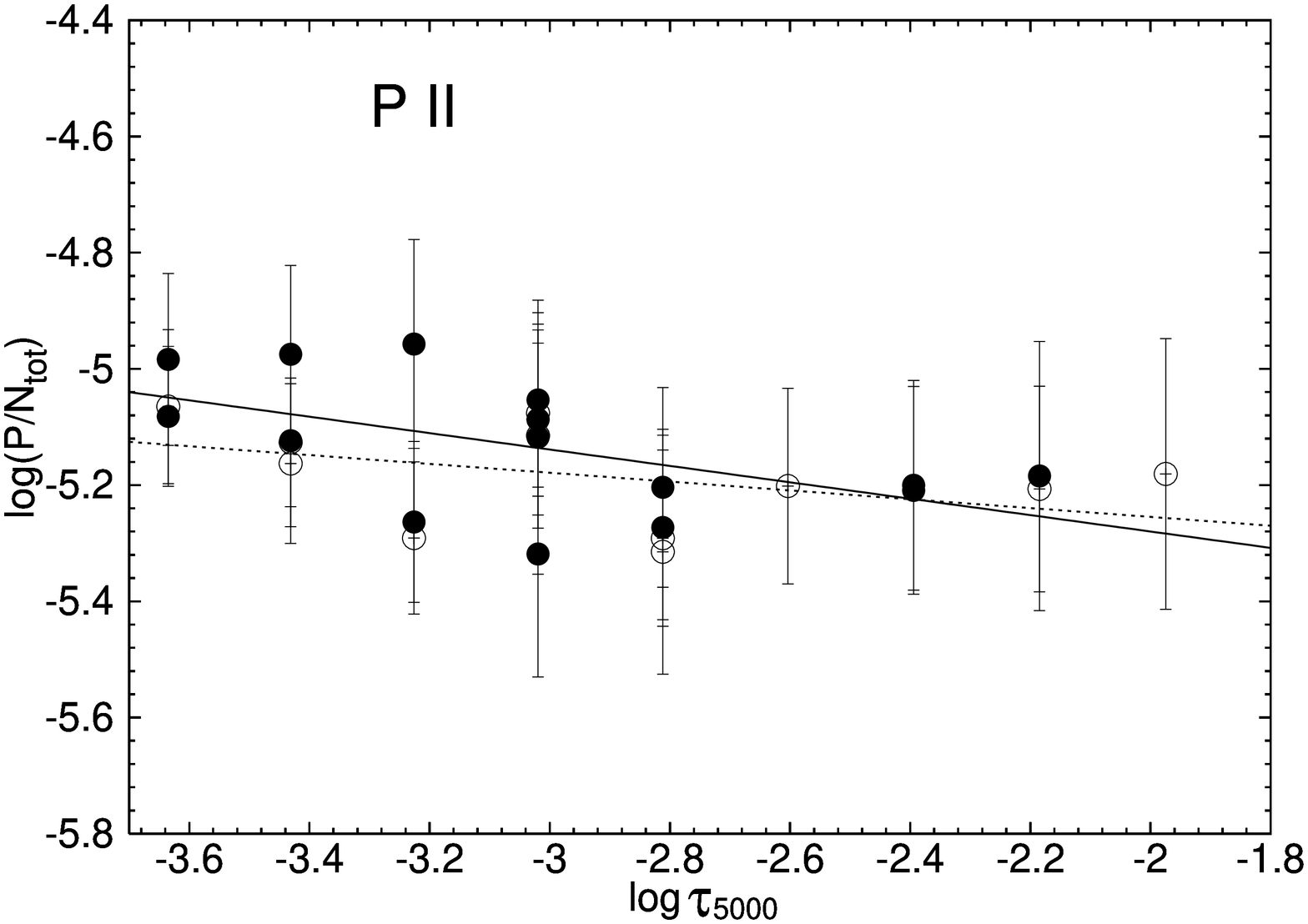}
\end{tabular}
}
\caption{The same as in Fig.~\ref{WF4-3085}, but for lines extracted from WF2-2541 spectrum.
\label{WF2-2541}}
}
\end{figure*}

\begin{table*}[th]
\parbox[t]{\textwidth}{
\centering
\caption[]{Slopes of vertical abundance stratification calculated
for the linear least-square fit to the data assuming zero microturbulence.}
\begin{tabular}{lrrrrrr}
\hline
\hline
Cluster/Star&\multicolumn{2}{c}{Slopes of $\log(Fe/N_{tot})$ vs.}& n &
             \multicolumn{2}{c}{Slopes of $\log(P/N_{tot})$ vs.}& n \\
            &\multicolumn{1}{c}{$\log{E_{i}}$}&\multicolumn{1}{c}{$\log{\tau_{5000}}$}& &
             \multicolumn{1}{c}{$\log{E_{i}}$}&\multicolumn{1}{c}{$\log{\tau_{5000}}$}& \\
\hline
M13/WF2-2541&0.18$\pm$0.04&0.17$\pm$0.03& 53&-0.67$\pm$0.31&-0.14$\pm$0.06&16\\
M13/WF4-3085&0.10$\pm$0.03&0.15$\pm$0.02&117&0.54$\pm$0.31&0.24$\pm$0.04&12\\
M13/WF4-3485&-0.06$\pm$0.03&-0.02$\pm$0.03&50&\multicolumn{1}{c}{...}&\multicolumn{1}{c}{...}& ...\\
M15/B84     &0.04$\pm$0.05&0.03$\pm$0.04& 25&\multicolumn{1}{c}{...}&\multicolumn{1}{c}{...}& ...\\
             &      &  &        &\multicolumn{2}{c}{Slopes of $\log(Ti/N_{tot})$ vs.} & \\
M15/B267    &0.49$\pm$0.06&0.33$\pm$0.03& 46&0.47$\pm$0.10&0.36$\pm$0.07&16\\
M15/B279    &0.23$\pm$0.04&0.23$\pm$0.02& 71&0.21$\pm$0.11&0.30$\pm$0.08&11\\
\hline
\end{tabular}
\label{slope}
}
\end{table*}

\section{Vertical abundance stratification}
\label{vert}

Applying the technique described in Subsec.~\ref{proc}, we have attempted to determine if the iron
abundance is vertically stratified in the atmospheres of the six selected BHB stars. For comparison,
we have also investigated the possible vertical stratification of titanium and phosphorus for some
BHB stars, where these elements are represented by a sufficient number of spectral lines.

All of the selected BHB stars are slowly rotating and have sharp absorption lines that result in
only 8 to 13 spectral bins per line profile. This fact, together with the S/N and the
uncertainties in the atomic data, are the main contributors to errors in the abundance inferred
from a single line. This makes the detection of weak vertical abundance variations difficult.

\begin{figure*}[th]
\parbox[t]{\textwidth}{
\centerline{%
\begin{tabular}{@{\hspace{+0.05in}}c@{\hspace{+0.1in}}c}
a) & b)\\
\includegraphics[angle=-90,width=3.5in]{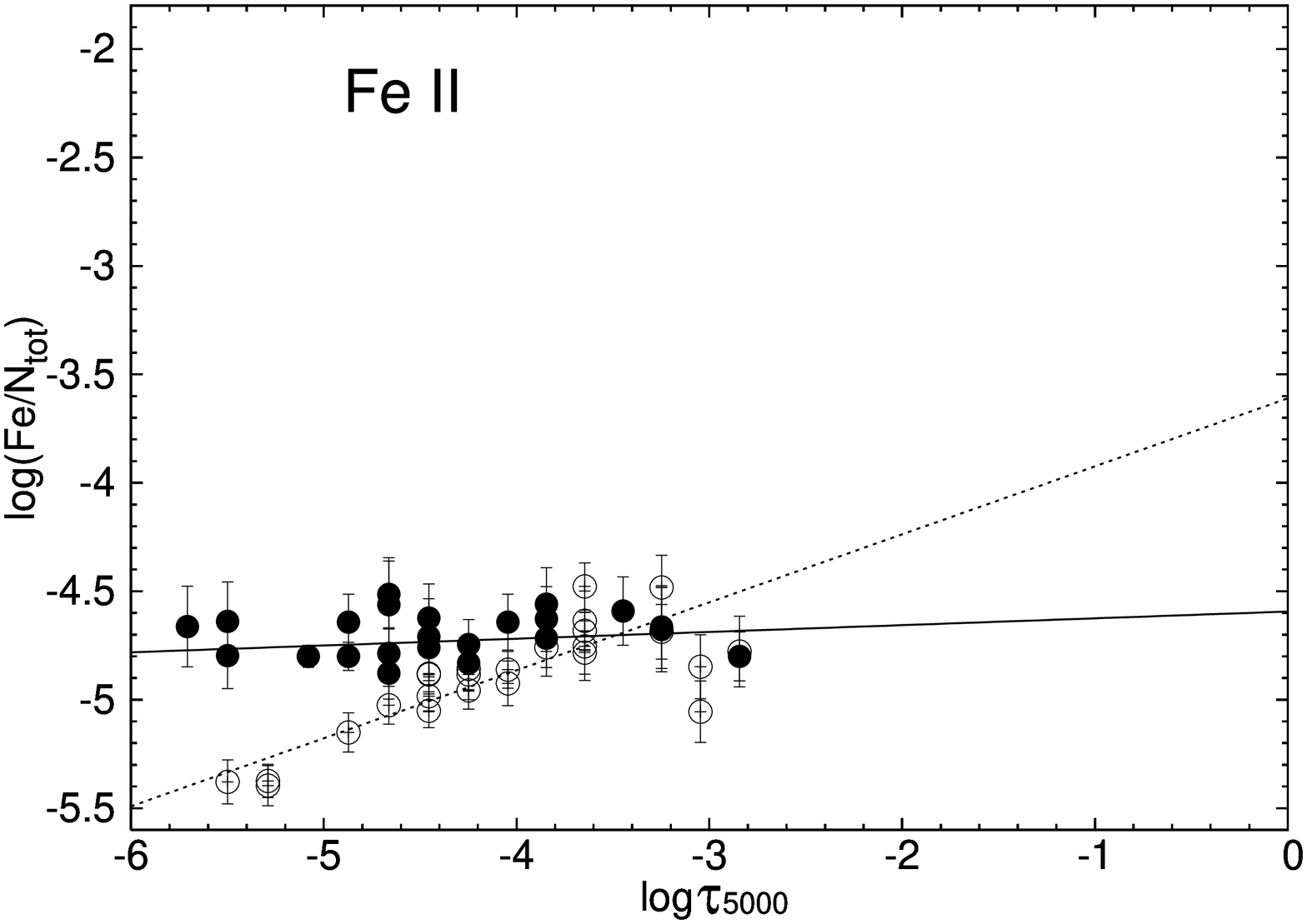} &
\includegraphics[angle=-90,width=3.5in]{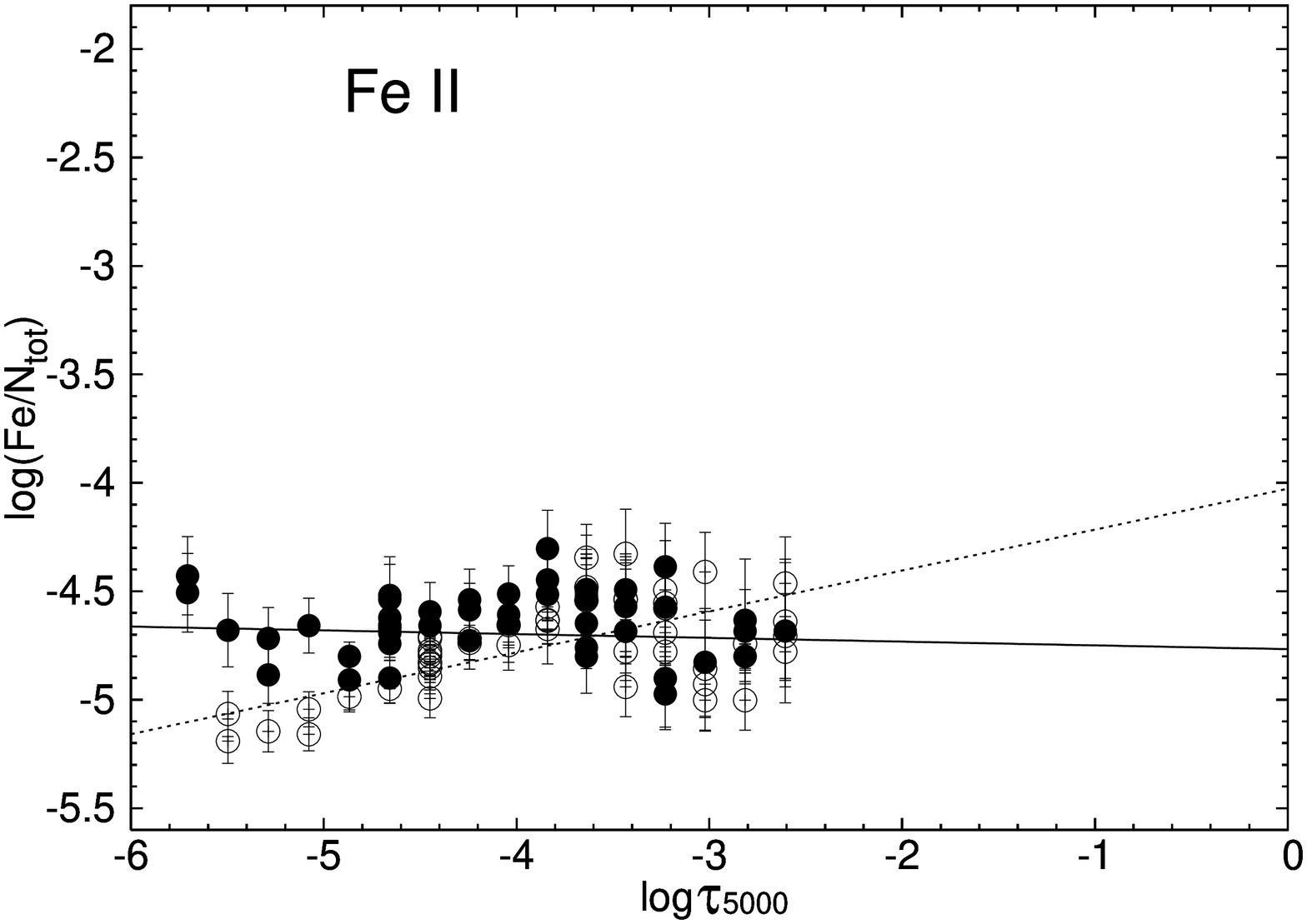}
\end{tabular}
}
\caption{The same as in Fig.~\ref{B267}, but for Fe\,{\sc ii} lines extracted for (a) B84 and (b) WF4-3485.
\label{two}}
}
\end{figure*}

Our abundance analysis for $\xi = 0$ km\,s$^{-1}$ shows that in the atmospheres of four BHB
stars (B267, B279, WF4-3085 and WF2-2541) the iron abundance generally increases towards the
deeper atmospheric depths (see Figs.~\ref{B267} to \ref{WF2-2541}). It should be noted that the
range of optical depths diagnosed by the iron lines is much smaller in the stars B279 and WF2-2541
than in the other two stars shown here. To estimate the stratification
profile of the iron abundance we fit the data with a straight line using a least-square
algorithm. We have also checked the dependence of each element's abundance with respect to the lower level
excitation potential $E_{i}$ for the sample of analyzed lines (see as an example Fig.~\ref{excitation}).
Slopes of the linearly approximated dependence of each element's abundance with respect to $\log{E_{i}}$
and $\log{\tau_{5000}}$ are given in Table~\ref{slope} for the case of zero
microturbulent velocity. Table~\ref{slope} shows a clear correlation between the slope of
the abundance versus $\log{\tau_{5000}}$ and $\log{E_{i}}$ for the elements
considered in each star.
This correlation strongly suggests that the obtained stratification is physical
and not due to 
uncertainties in the description of the temperature profile of the stellar atmosphere.

The detected stratification of iron, obtained in some of the stars studied here,
is far too large ($>$ 2 dex) to be
interpreted as the result of measurement errors. Our stellar atmosphere models were calculated taking
into account the enhanced iron abundance and helium depletion (Behr~\cite{Behr03b}) and the error in
estimating the iron abundances, for a given model atmosphere,
is expected to be less than 0.2~dex. Meanwhile, Khan~\&~Shulyak~(\cite{K+S07}) have
shown that using atmospheric models with varying Fe abundance (from one to ten times solar) can
modify the abundances inferred from observed line profiles by up to $\pm$0.25~dex.

From the results in Table~\ref{slope} we can conclude that no detectable iron stratification exists in WF4-3485 and B84.
Two stars (B267 and B279) show quite large slopes of iron abundance with respect to both $\log{E_{i}}$ and
$\log{\tau_{5000}}$. The corresponding slopes for the stars WF2-2541 and WF2-3085 are smaller but still
statistically significant. The least significant of these slopes is that of the
abundance versus $\log{E_{i}}$ for WF2-3085 which has only approximately a $3\sigma$ value.

Another factor that must be evaluated that could mimic stratification is an error
in the effective temperature of the underlying atmospheric model used. To verify
the potential importance of this factor, we calculated the Fe stratification with models
with an effective temperature higher
and lower by 1000 K from the $T_{\rm eff}$ values listed in Table~\ref{tab2}. We also calculated
the stratification with models assuming solar He and Fe abundances to evaluate the
effect of this change on the inferred stratification profiles. Our simulations show that models with higher
$T_{\rm eff}$ and models with solar abundances usually
slightly decrease the slope of the abundance with respect to $\log{E_{i}}$,
while models with lower $T_{\rm eff}$ increase this slope for all elements.
In the case
of iron in WF4-3085, the model with higher $T_{\rm eff}$ and
the model with solar abundances result in a negligible slope of its abundance versus
$\log{E_{i}}$. Therefore, the results are not sufficient
to report unambiguous detection of vertical stratification of iron in WF4-3085.
The slopes for iron found for the other three stars (B267, B279 and WF-2541) are
still significant in all these models, and stratification is therefore confirmed.

The two BHB stars (B84 in M15 and WF4-3485 in M13) which do not show any signs of
stratification of their iron abundances are not different from the other stars
with regards to their stellar atmosphere characteristics and rotation (see Table~\ref{tab2}).
However, these two stars possess very small titanium and phosphorus abundances as compared to the other
stars studied here. Their spectra have very weak or invisible Ti\,{\sc ii} and P\,{\sc ii} lines.
Their iron abundance is also close to the solar value, at least for
$\xi=0$\,km\,s$^{-1}$ (see Table~\ref{tab5}).

We observe an upturn in the iron abundance obtained assuming $\xi=0$\,km\,s$^{-1}$
at low optical depths for some of the stars studied
(e.g. Figs.~\ref{B267}a,~\ref{B279}a,~\ref{WF4-3085}a,~\ref{WF2-2541}a).
The points on the figures corresponding to the upturn feature are the strong Fe\,{\sc ii} lines
and are not taken into account during the least-square fit of the data. 
To estimate the influence of microturbulence we have performed an abundance analysis with $\xi=$ 2\,km\,s$^{-1}$,
which leads to elimination of the upturn (see, for example, Fig.~\ref{B267}a).
Performing a set of simulations with different microturbulent velocities, we have determined the minimum value
of the microturbulence ($\xi_{min}$) for which the upturn disappears for each star studied. These values are
reported in Table~\ref{tab5}.
As the inclusion of microturbulent velocity amplifies the iron stratification,
the stratification obtained for iron in the first four stars studied with $\xi=$ 0\,km\,s$^{-1}$ can be considered
to be a lower limit to the possible stratification in these stars.
A more detailed study including abundance stratification
in the calculation of the synthesized spectra might shed some light on this strange behaviour,
but this is outside of the scope of the present paper.

\begin{table*}[th]
\parbox[t]{\textwidth}{
\centering
\caption[]{Characteristics of our sample of hot BHB stars obtained from spectral simulations.}
\begin{tabular}{lcccccrccc}
\hline
\hline
Cluster/Star& v$\sin{i}$&$V_{\rm r}$&$\xi_{min}$&\multicolumn{3}{c}{[Fe/H]}&
\multicolumn{3}{c}{[P/H]} \\
            &km s$^{-1}$&km s$^{-1}$&km s$^{-1}$&$\xi$=0 km s$^{-1}$&$\xi$=2 km s$^{-1}$& n &$\xi$=0 km s$^{-1}$&$\xi$=2 km s$^{-1}$& n \\
\hline
M13/WF2-2541&2.4$\pm$0.5&-257.3$\pm$0.5&0.0+0.3    &-0.12$\pm$0.20&-0.19$\pm$0.26& 53&+1.51$\pm$0.11&+1.46$\pm$0.10&16\\
M13/WF4-3085&3.4$\pm$0.8&-255.4$\pm$0.5&0.5$\pm$0.3&+0.72$\pm$0.29&+0.59$\pm$0.37&117&+1.84$\pm$0.13&+1.84$\pm$0.15&12\\
M13/WF4-3485&3.0$\pm$0.9&-246.9$\pm$0.9&1.5$\pm$0.3&-0.04$\pm$0.50&-0.17$\pm$0.27&50&\multicolumn{1}{c}{...}&
\multicolumn{1}{c}{...}& ... \\
M15/B84     &4.0$\pm$1.3&-108.3$\pm$0.4&0.0+0.3    &-0.14$\pm$0.10&-0.35$\pm$0.24&25&\multicolumn{1}{c}{...}&
\multicolumn{1}{c}{...}& ... \\
            &         &            &      &  &        & &\multicolumn{3}{c}{[Ti/H]} \\
M15/B267    &6.2$\pm$1.2&-114.7$\pm$0.7&0.5$\pm$0.3&+0.45$\pm$0.37&+0.21$\pm$0.51& 46&+0.09$\pm$0.20&-0.08$\pm$0.24&16\\
M15/B279    &4.7$\pm$1.6&-104.5$\pm$0.6&1.0$\pm$0.3&+0.42$\pm$0.34&+0.20$\pm$0.40& 71&-0.44$\pm$0.10&-0.54$\pm$0.11&11\\

\hline
\end{tabular}
\label{tab5}
}
\end{table*}

Figure~\ref{B267}b indicates that B267 shows a possible signature of vertical stratification of
the titanium abundance.
The respective slopes of Ti abundance versus $\log{E_{i}}$ and
$\log{\tau_{5000}}$ are significantly higher than zero for this star and are
statistically significant (see Table~\ref{slope}).
For B279, we can not confidently conclude that stratification exists since the
slope of the abundance with respect to $\log{E_{i}}$ becomes weak when
using either solar abundances or assuming a $T_{\rm eff}$ increase of 1000 K. The
slope of Ti abundance with respect to $\log{\tau_{5000}}$ also becomes weak
in the model assuming a $T_{\rm eff}$ increase of 1000 K. It should be noted
that the Ti lines in B279 do not sample a large portion of the atmosphere and
this fact renders the detection of stratification more difficult.

It is clear that phosphorous shows no clear signs of stratification in WF2-2541.
For WF4-3085 we cannot make firm conclusions concerning phosphorous stratification
since its abundance variation in the
range of optical depths under consideration is only approximately 0.4 dex. Also,
the slope of its abundance with respect to $\log{E_{i}}$ is not statistically
significant.

\section{Mean abundances and velocities}
\label{mean}

The mean photospheric abundances derived for the selected BHB stars from an analysis of the
available iron, titanium and phosphorus lines are reported in Table~\ref{tab5}. 
For each chemical element, three columns are shown containing the mean abundance for simulations
with microturbulent velocity $\xi$= 0 and 2 km s$^{-1}$ respectively, and the number of analyzed line profiles.
The reported uncertainties are equal to the standard deviation calculated from the results of individual line
simulations for all lines considered.

The derived heliocentric radial and projected rotation velocities as well as microturbulent velocities
(see Table~\ref{tab5}) are generally in agreement with the data provided by Behr (\cite{Behr03b})
for these stars. Some inconsistencies are found between $v\sin{i}$ values obtained for
WF2-2541 ($v\sin{i}=0.0^{+4.07}_{-0.0}$ km s$^{-1}$) and B279 ($v\sin{i}=5.92^{+1.6}_{-1.69}$ km s$^{-1}$),
but they are within the respective error bars given by Behr (\cite{Behr03b}).  The precision of our velocity
estimates appears to be comparatively higher because of a more stringent selection of analysed line profiles.
Our measurements of the average abundance of iron, titanium and phosphorus also agree with the corresponding values
published by Behr (\cite{Behr03b}), taking into account the error bars and the value of the applied
microturbulent velocity.
Small differences between our results and abundances from the aforementioned paper are
not surprising. Our average abundance is calculated with the individual abundances obtained from each line,
while Behr (\cite{Behr03b}) fitted the whole spectrum of each element to obtain its abundance.

\section{Discussion}
\label{discuss}

In this paper we continue our attempts to detect vertical abundance stratification in
the atmospheres of BHB stars. After the report by Bonifacio et al. (\cite{Bonifacio+95})
of the vertical stratification of helium in the atmosphere of Feige 86, it became clear
that the abundances of other chemical species may also be stratified.   We devoted special interest
to iron because hot BHB stars usually have an enhanced iron abundance (e.g. Behr~\cite{Behr03b}),
suggesting that this element may be strongly affected by diffusion.
Analysing the spectra of another hot BHB star HD~135485 (Khalack et al. \cite{Khalack+07}) we did not
find direct evidence of iron stratification, but revealed strong signatures of
sulfur depletion in the deeper atmospheric layers.  However, HD~135485
is different from the other BHB stars in that its spectrum shows evidence of helium enrichment
(in comparison with the solar abundance), while in the atmospheres of the other BHB stars
helium is depleted.

Therefore, we directed our attention to BHB stars where the iron abundance is near the
solar abundance or enhanced, and helium is depleted.
The results obtained argue that at least three stars (B267 and B279 in M15 and 
WF2-2541 in M13) show clear signatures of vertical stratification of their iron abundance,
while for WF4-3085 the results are suggestive, but not conclusive.
The other two stars studied here (B84 in M15 and WF4-3485 in M13) do not show stratification of iron
and their averaged iron abundance is close to solar (but is enhanced in comparison with its cluster value).
B267 shows also a signature of vertical stratification of titanium (see Fig.~\ref{B267}b).

Since our simulations show that the turnup feature observed in the iron stratification profile
is strongly dependent on microturbulent velocity, the value of the abundance at low optical depths
is uncertain. Of course, if the theoretical framework supposes that these abundance gradients are due to atomic
diffusion, microturbulence should be weak since a stable atmosphere is needed for diffusion to be dominant.
It should be noted that for corresponding optical depths the abundance profile of iron is similar in the
three BHB stars that exhibit stratification. The reason that the other two stars in our study do not show
clear signs of iron stratification (B84 and WF4-3485) might be related to evolutionary effects or the presence of other
competing hydrodynamical processes. The absence of Ti\,{\sc ii}
and P\,{\sc ii} lines in their spectra
might be evidence of this.

In conclusion, the results shown here add to the mounting evidence of the existence of vertical abundance
stratification, and hence atomic diffusion, in the atmospheres of BHB stars.

\begin{acknowledgements}
This research was partially funded by the Natural Sciences and Engineering Research Council
of Canada (NSERC). We thank the R\'eseau qu\'eb\'ecois
de calcul de haute performance (RQCHP) for computing resources.
GAW acknowledges support from the
Academic Research Programme (ARP) of the Department of National Defence (Canada).
BBB thanks all the dedicated people involved in the construction and
operation of the Keck telescopes and HIRES spectrograph. He is also
grateful to Judy Cohen, Jim McCarthy, George Djorgovski, and Pat C\^ot\'e
for their contributions of Keck observing time.
We are grateful to Dr. T.Ryabchikova and Dr. L.Mashonkina
for helpful discussions and suggestions.

\end{acknowledgements}

\end{document}